\documentclass[twocolumn,nofootinbib,tightenlines,nobibnotes, amssymb,amsmath,aps,prd,reprint, superscriptaddress,preprintnumbers]{revtex4-1}
\usepackage{graphicx}
\usepackage{epsfig}
\usepackage{amssymb,amsmath,amsfonts,hyperref}
\usepackage{wasysym}
\usepackage{latexsym}
\usepackage{eucal}
\usepackage{verbatim}
\usepackage[normalem]{ulem}
\usepackage{float}

\usepackage[draft,color,final]{showkeys}
\definecolor{refkey}{rgb}{0.9451,0.2706,0.4941}
\definecolor{labelkey}{rgb}{0.9451,0.2706,0.4941}

\allowdisplaybreaks
\allowbreak

\newcommand{\f}{\frac}

\newcommand{\bea}{\begin{eqnarray}}
\newcommand{\eea}{\end{eqnarray}}
\newcommand{\ba}{\begin{align}}
\newcommand{\ea}{\end{align}}

\newcommand{\md}{\mathcal{D}}

\newcommand{\lsim}{\mathrel{\hbox{\rlap{\lower .55ex
\hbox{$\sim$}} \kern-.3em \raise.4ex \hbox{$<$}}}}
\newcommand{\gsim}{\mathrel{\hbox{\rlap{\lower.55ex
\hbox{$\sim$}} \kern-.3em \raise.4ex \hbox{$<$}}}}

\def\be{\begin{equation}}
\def\ee{\end{equation}}
\def\bal{\begin{array}{l}}
\def\ba#1{\begin{array}{#1}}  
\def\ea{\end{array}}
\def\bea{\begin{eqnarray}}
\def\eea{\end{eqnarray}}
\def\beas{\begin{eqnarray*}}
\def\eeas{\end{eqnarray*}}

\def\nn{\nonumber}

\def\l{\lambda}

\def\lan{\langle}
\def\ran{\rangle}
\def\f{\frac}

\def\bit{\begin{item}}
\def\eit{\end{item}}
\def\benu{\begin{enumerate}}
\def\eenu{\end{enumerate}}

\def\tr{{\rm tr}}

\begin{document}

\title{Spectral Form Factor in Non-Gaussian Random Matrix Theories}
\author{Adwait Gaikwad}
\email{adwait@theory.tifr.res.in}
\affiliation{\small{Tata Institute of Fundamental Research, 400005 Mumbai, India.}}
\author{Ritam Sinha}
\email{ritamsinha.physics@gmail.com}
\affiliation{\small{Tata Institute of Fundamental Research, 400005 Mumbai, India.}}
\affiliation{\small{Instituto de F\'{i}sica Te\`{o}rica, UAM-CSIC, 28049 Madrid, Spain.}}

\preprint{TIFR/TH/17-24}

\begin{abstract}
We consider Random Matrix Theories with non-Gaussian potentials that have a rich phase structure in the
large $N$ limit. We calculate the Spectral Form Factor (SFF) in such models and present them as interesting examples of dynamical models that display multi-criticality at short time-scales and universality at large time scales. The models with quartic and sextic potentials
are explicitly worked out. The disconnected part of the Spectral Form Factor (SFF) shows a change in its decay behavior exactly at the critical points of each model. The dip-time of the SFF is estimated in each of these models.
The late time behavior of all polynomial potential matrix models is shown to display a certain universality. This is related to the universality in the short distance correlations of the mean-level densities. We speculate on the implications of such universality for chaotic quantum systems including the SYK model. 
\end{abstract}

\maketitle
\section{Introduction}
Quantum chaos in many body systems has attracted the attention of physicists across various disciplines for over many decades now. Given that classically chaotic systems could be characterised using only a few stringent conditions, one hoped that the same would be true for chaotic quantum systems. However, it turns out that chaos in many body quantum systems is a much more complicated phenomenon. A key role to characterise chaos in quantum systems is played by Random Matrix Theories (RMTs). The idea (first proposed in \cite{Dyson:1962es1}) is to look at the energy spectrum of a given quantum system. Typically, for a few body system, if the levels follow the \emph{Wigner's semi-circle law}, then such systems are very well approximated by RMTs. For many body quantum systems, a sharp diagnostic of chaos is obtained by looking at the distribution of the nearest neighbour spacings (NNSD) of the energy levels of the system. The system is termed chaotic if the distribution is of the \emph{Wigner-Dyson} type \cite{PhysRevLett.52.1}\footnote{See \cite{Garcia-Garcia:2016mno} for a discussion of the SYK model.}.
Such a behaviour is considered to be the hallmark of chaos in quantum many-body systems. Interestingly, the above expression has been shown to hold analytically only for few-body Hamiltonians ($N$ finite). For many-body quantum systems, the result has been established only numerically. Nonetheless, the remarkable validity of this result over a wide range of systems leads one to believe that chaos at the quantum level is an extremely rich phenomenon, that may exist irrespective of whether the system has a classical counterpart.

That being said, classically chaotic systems can still teach us a thing or two about quantum chaos. 
An important diagnosis for chaos in classical systems is the sensitive dependence on initial conditions, characterised by a positive Lyapunov exponent. An analog of this definition for a few-body quantum system was outlined in \cite{Larkin:1969} using out-of-time-ordered correlators (OTOCs). Recently, the OTOCs have been widely adopted for the analysis of chaos in quantum many body systems, including large $N$ CFTs and even Black Holes \cite{Kitaev, Maldacena:2015waa, Shenker:2013pqa, Roberts:2014ifa}. 

The novelty of OTOCs is in that it helps us explore an interesting timescale, called \emph{scrambling  time}, in systems with a large number of degrees of freedom. First discussed in the context of black holes in \cite{Hayden:2007cs, Sekino:2008he},
this timescale characterises how long it effectively takes for an initial localized perturbation to spread over all degrees of freedom in a system. Usually, for chaotic many body systems, such a timescale is parametrically larger than the thermalisation timescale in the system. In order to make any connection between the two afore-mentioned definitions of quantum chaos, it would be worthwhile to understand whether and how scrambling of degrees of freedom ultimately leads to the onset of RMT behaviour in the system.
In fact, some concrete steps along this direction have already been undertaken in the context of the SYK model. The SYK model was first shown to be chaotic using OTOCs in \cite{Kitaev,Maldacena:2016hyu}. More recently, the model was shown to have a late-time behaviour similar to Gaussian RMTs \cite{Cotler:2016fpe}. However, the analysis in \cite{Cotler:2016fpe} was mostly numerical, and a more analytic understanding of this cross-over remains an open problem.

An intriguing aspect of RMTs is an inherent timescale, called the \emph{ramp} time \cite{Gharibyan:2018jrp}. It is the timescale below which no universal correlations between the energy levels exists. It usually indicates the onset of Random Matrix behaviour in the system. 
Above the \emph{ramp} time, the universal correlations between the nearby energy levels starts to manifest themselves. A good diagnostic for exploring such timescales is an observable called the Spectral Form Factor (SFF). The SFF was introduced long ago \cite{Brezin:1997} and recently re-introduced \cite{Garcia-Garcia:2016mno, Dyer:2016pou, Cotler:2016fpe, Krishnan:2016bvg, Balasubramanian:2016ids, delCampo:2017bzr, Krishnan:2017ztz, Mehta, Papadodimas:2015xma} as a tool for probing the spectra of quantum systems. It is defined as,
\begin{equation}
|Z(\beta + i t)|^2=\sum_{m,n}e^{-\beta(E_m + E_n)}e^{-i t(E_m-E_n)}
\end{equation}
where $Z(\beta)$ is the partition function of a quantum system and $\beta$ is the inverse temperature. 
For $\beta=0$, it is easy to see that the above expression picks out contributions only from the differences between the 
nearest neighbour energy eigenvalues at very late times. This makes it a good probe for understanding not only the discreteness 
of the energy spectrum, but also gives a way to characterise a chaotic system based on its NNSD.
In fact, the SFF (when averaged over Gaussian Random Matrices) has a very particular behavior at large $N$ characterised by an intial decay, followed by a linear rise and saturation. This indicates how the SFF perceives the nearest neighbour energy spectrum. In fact in \cite{Cotler:2016fpe}, the SFF was used to show the similarities between the late time behaviour of the SYK model and Gaussian RMTs.

The goal of this paper is to understand whether and how the structure of the SFF is modified under a controlled deformation of the RMT from a Gaussian to a non-Gaussian model. For this, we will consider the SFF corresponding to polynomial non-Gaussian matrix models, that have a richer phase structure as compared to the Gaussian model\footnote{Phase transitions, corresponding to such a phase structure, were first shown to exist in Unitary Matrix Models and are termed the Gross-Witten-Wadia (GWW) phase transitions \cite{Wadia:1980cp, Gross:1980he}.}. Simultaneously, we will try to understand how such deformations affect the chaotic structure of the energy eigenvalues. This shall be reflected in any changes in the onset time for the ramp in the SFF.
We specifically work out the Quartic and the Sextic models, where 
we average the SFF over each of these non-Gaussian ensembles and make the following observations,
\begin{enumerate}
 \item The initial decay behavior is the same as for the Gaussian ensemble, when the non-Gaussian ensemble is away from its critical points.
 \item At the critical points, the decay behavior of the SFF changes over to a faster decay, displaying a different power law behavior. This gives an explicit example where the multi-critical behavior at equilibrium is extended to non-equilibrium dynamics.
 \item The \emph{dip-time} estimate changes with the change in the fall-off behavior at criticality, providing an upper bound on the \emph{ramp-time}. This in turn provides us with an idea of when RMT behaviour sets into the system.
 \item The late time behavior of the SFF continues to be characterised by a linear growth followed by saturation, indicating no change in the NNSD.
\end{enumerate}

We finish by speculating on how to obtain a possible analytic understanding of the crossover between the two different descriptions of chaotic behaviour in the same system.

This paper is organized as follows. 
In \ref{sec-2}, we start with describing the Gaussian Matrix Model and calculate its mean level density in the large $N$ limit. That is followed by a description of two specific non-Gaussian Matrix Models, namely the quartic and the sextic models. We then proceed to calculate the mean-level densities for these models. In \ref{sec-3}, we introduce the Spectral Form Factor and the method for calculating it in the large $N$ limit, by separating it out into a connected and a disconnected part. 
In this section, we state our main results regarding the change in the decay behavior of the disconnected part of the SFF exactly at the critical points of these models. In \ref{sec-4} we estimate the dip-time (indicating the onset of ramp) for these models.
In \ref{sec-5}, we talk about the implications of non-Gaussian Matrix Models on chaotic systems.

\section{Matrix Models: Gaussian and Non-Gaussian\label{sec-2}}
The approach to the complete non-perturbative solution of RMTs with the most general polynomial potential (at large $N$) is well established\cite{Mehta}\footnote{See Appendix A and B for a comprehensive review of the solution.}. Equipped with this solution, we shall now consider specific forms of the potential and study the corresponding matrix models case by case to understand their properties. The idea would be to understand their implications in describing chaotic behavior, in particular, the spectral form factor, which we shall define in the next section.

\subsection{Gaussian Matrix Model}

\begin{figure}[h]
\centering
 \includegraphics[width=220pt, height=140pt]{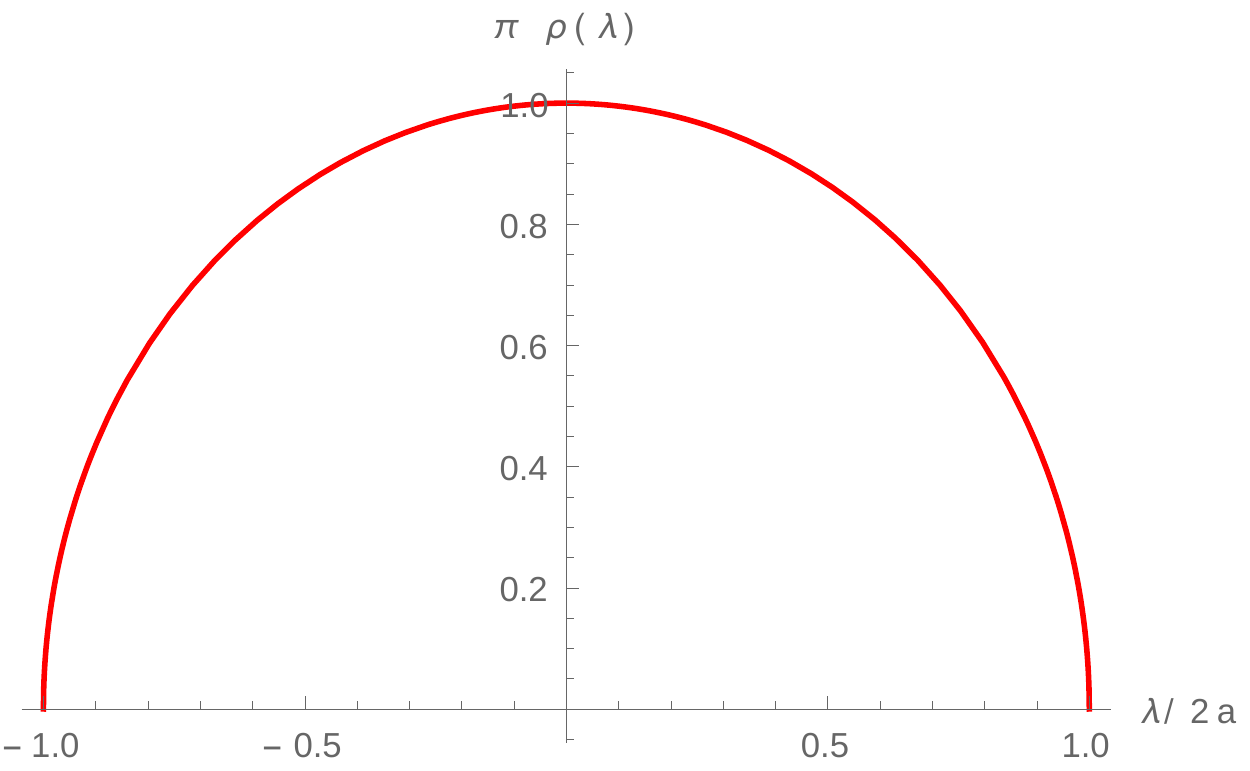}
 \caption{\footnotesize{Wigner's semi-circle law}}
 \label{fig-1}
\end{figure}
We begin by considering a Gaussian potential: $V(M)=\f12 M^2$. The corresponding action in the continuum limit is,
\begin{widetext}
\begin{equation}
 \mathcal{Z} = \left(\prod_{i=1}^{N} \int d\lambda_i \right) 
 \exp\left[{-\frac{N^2}{2}\left(\int\ d\lambda\ \rho(\lambda)\ \lambda^2 
 - 2 \int d\lambda\ d\mu\ \rho(\lambda)\  \rho(\mu) \log(|\lambda-\mu|)\right)}\right]
 \label{GUE}
\end{equation}
\end{widetext}
Extremising this action with respect to the level density $\rho(\lambda)$ gives us,
\begin{equation}
\f12\lambda^2=2\int_{\mathbb{R}}\,d\mu \rho(\mu)\log|\lambda-\mu|
\end{equation}
Taking a derivative of the overall equation w.r.t. $\lambda$ gives \eqref{saddle-pt} with $V'(\lambda)=\lambda$.
For a Gaussian potential in a symmetric interval $\big(-1,1\big)$, the resolvent takes the 
following form,
\begin{equation}
\omega(\lambda\pm i 0)= \f{\lambda}2 \pm \f12 \sqrt{\lambda^2-4}
\end{equation}
From the properties of the resolvent (stated in Appendix \ref{B}), it follows that the level density takes the unique form,
\begin{equation}
\rho(\lambda)=\f1{2\pi}\sqrt{4-\lambda^2}
\label{density-gauss}
\end{equation}
This is the \emph{Wigner's semi-circle law} for Gaussian matrices (shown in Fig. \ref{fig-1}) in the large $N$ limit.

\subsection{Non-Gaussian Matrix Models}
\subsection*{Quartic Potential}
\begin{figure}[h]
\centering
 \includegraphics[width=220pt, height=140pt]{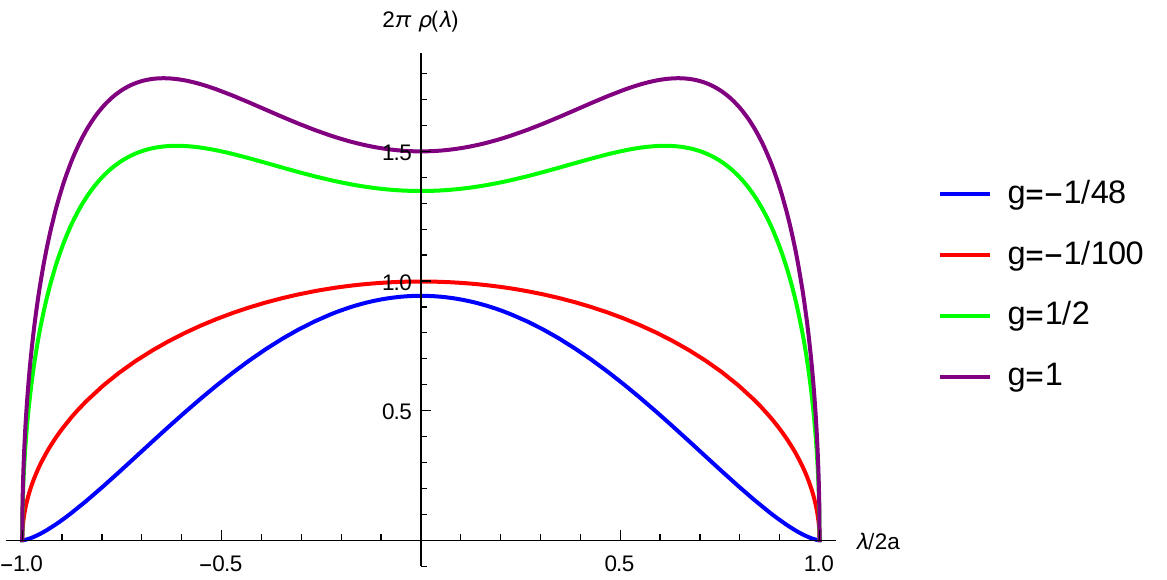}
 \caption{\footnotesize{ Mean-level density of the quartic model at $g=\{-1/48,-1/100,1/2,1\}$}}
 \label{fig-21}
\end{figure}
As the simplest example of a non-Gaussian matrix model, we consider a quartic potential of the form,
\begin{equation}
V(M)=\f{1}{2}  M^2 + \f{g}{N} M^4
\end{equation}
where, in the second term, $g$ is the coupling constant and the power of $N$ is so adjusted as to make the quartic interaction strength comparable to the quadratic term in the action at large $N$.
This model has been shown (\cite{Brezin:1977sv, Bhanot:1990bd, Mandal:1989ry}) to have a 
critical point at $g=g_c=-1/48$, below which it is supposed to have no solutions. At $g_c$ 
the free-energy of the model has a non-analyticity. It is our goal to study the behavior of the spectral form factor computed using one-cut solution\footnote{An $n$-cut solution is a function which has support on $n$ disjoint intervals. Throughout this paper, we will work with one cut solutions only.} and see how it changes at the critical point. 
It can be argued (Appendix \ref{B}) that there is a unique resolvent corresponding to the saddle point equation in this phase. 
This unique form of the resolvent is,
\begin{equation}
\omega(\lambda)=\f\lambda2 + 2 g \lambda^3 -\left(\f12 + 4 g a^2 + 2 g \lambda^2 \right)\sqrt{\lambda^2-4 a^2}
\label{resolvent}
\end{equation}
where $a$ is defined through the constraint,
\begin{equation}
12ga^4 + a^2 -1 =0
\label{constraint}
\end{equation}
The level density is found from (\ref{resolvent}) to be,
\begin{equation}
\rho(\lambda)=\f1{\pi}\left(\f12 + 4 g a^2 + 2 g \lambda^2 \right)\sqrt{4a^2-\lambda^2}
\label{density-quartic}
\end{equation}
The end-points of the interval $(-2a,2a)$ (on which $\rho(\lambda)$ is real) depend on the value of the coupling constant 
$g$ through (\ref{constraint}) as,
\begin{equation}
a^2=\f1{24 g}(\sqrt{1+48 g}-1)
\end{equation}
It is easy to see from above that $a^2$ will have imaginary values for $g\le-1/48$ and hence, there are no real solutions of (\ref{constraint})
that exist below $g_c=-1/48$. \\
\noindent
One shortcoming of this analysis is the absence of two cut solutions, making
it impossible to access the phase below $g=g_c$. 
To remedy this, we need to go to a matrix model with a sextic potential \cite{Bhanot:1990bd}. Note that at $g=g_c$, the behaviour of the level density, near the edge, changes to (\ref{crit-quartic}) \footnote{We will consider the change in the edge behaviour of the level density as a sign of criticality.}.
\subsection*{Sextic Potential}
The form of the potential in the sextic model is,
\begin{equation}
V(M)=\f12 M^2 +\f{g}{N} M^4 + \f{h}{N^2}M^6
\end{equation}
The unique form of the resolvent for the one-cut solution with the above potential is given by,

\begin{widetext}
\begin{equation}
\omega(\lambda)=\f12 \lambda + 2 g \lambda^3 + 3 h \lambda ^5 - 
\bigg( 3 h \lambda^4 +(2g + 6h a ^2)\lambda^2+(\f12 + 4 g a^2 + 18 h a^4)\bigg)\sqrt{\lambda^2-4a^2} 
\label{sextic}
\end{equation}
\end{widetext}
with $h$ being defined through the constraint,
\begin{equation}
60 h a^6 + 12 g a^4 + a^2 -1=0
\label{constraint3}
\end{equation}
The level density, in the one-cut phase, corresponding to (\ref{sextic}) is,
\begin{equation}
\rho(\lambda)=\f1{\pi}\bigg( 3 h \lambda^4 +(2g + 6h a^2)\lambda^2 \hspace{3cm}\nonumber \\ \hspace{1.5cm}+ \left(\f12 + 4 g a^2 + 18 h a^4\right)\bigg)\sqrt{4a^2-\lambda^2}
\label{density-sextic}
\end{equation}
From (\ref{constraint3}) we see that the end points of the interval now depend on both the coupling constants $g$ and $h$.
To find the critical points in this model, we shall look for the points where the behavior of the level density changes near the edges of the spectrum. This calculation has been carried out in Appendix \ref{C}. In this model, there is a line of critical points in the $(g,h)$ plane, which starts from 
$(g,h)=(-1/48,0)$. The values of $g$ and $h$ on this critical line are \cite{Bhanot:1990bd},
\begin{equation}
g=\f1{12 a^4}(3-2a^2)\hspace{.5cm}h=\f1{60a^6}(a^2-2)
\label{g-h}
\end{equation}

 \begin{figure}[b]
 \includegraphics[width=220pt, height=140pt]{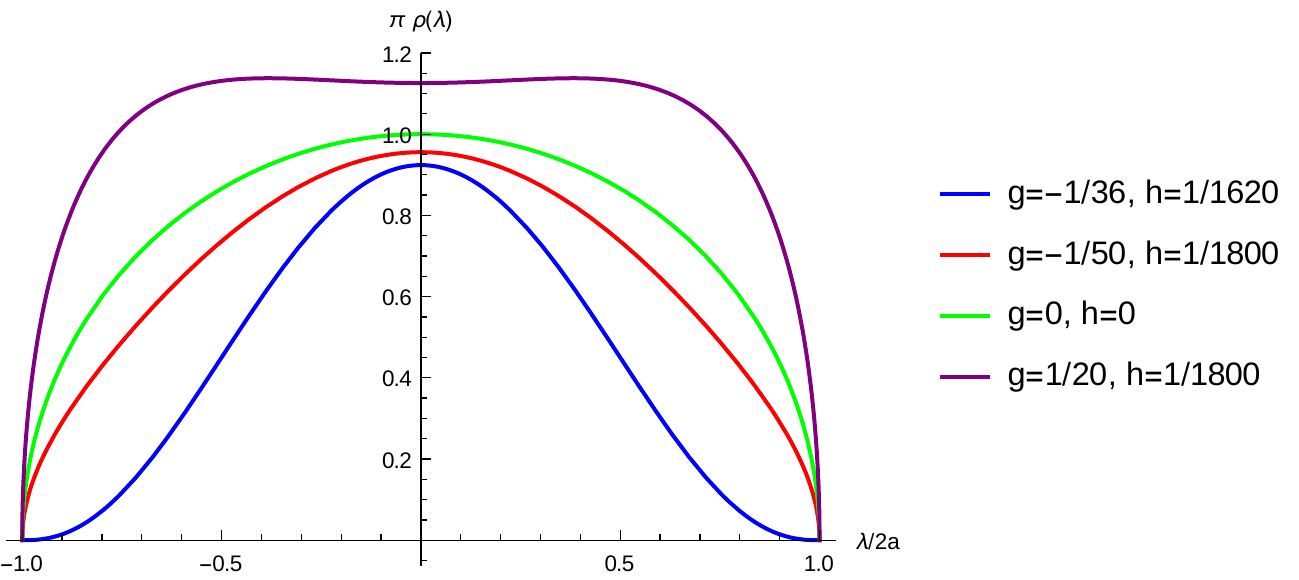}
 \caption{\footnotesize{Mean-level density for the sextic model at $(g,h)=\{(-1/36,1/1620),(-1/50,1/1800),(0,0),(1/20,1/1800)\}$ }}
\label{fig-22}
\end{figure}
\noindent
For $a^2=2$, the above values reduce to the critical point of the quartic model. The critical line encounters a tri-critical point for the following values of the couplings,
\begin{equation}
g=-1/36\hspace{.3cm}h=1/1620\hspace{.3cm}a^2=3
\label{tri-crit-pt}
\end{equation}

\noindent
Eliminating $a$ from \eqref{g-h}, we can construct a function $g_c\equiv g(h)$ along the critical line. Now, keeping $h$ fixed, if we start decreasing the value of $g$ starting from the origin, we find a number of solutions for the level density organised in the following way. For $g>g_c(h)$, and a fixed value of $h$, there exist all one, two and three-cut solutions. However, for $g<g_c(h)$, the one-cut solution vanishes, leaving behind the two and three-cut solutions which co-exist. From a naive analysis of the free-energy, it can be shown that in the region $g>g_c(h)$, the one-cut solution is the most stable, thereby dominating the thermodynamic behaviour. The nomenclature of the \emph{one-cut phase} for $g>g_c(h)$ and the \emph{two-cut phase} for $g<g_c(h)$ is, thereby derived from such a behvaiour. 

\section{Spectral Form Factor\label{sec-3}}
\noindent
The spectral form factor(SFF) was originally introduced as a probe for analysing the spectrum of quantum systems.
It is defined in terms of the analytically continued partition function as,
\begin{equation}
|Z(\beta + i t)|^2=\sum_{m,n}e^{-\beta(E_m + E_n)}e^{-i t(E_m-E_n)}
\end{equation}
From the R.H.S. of the above expression, it is easy to see that at very high temperatures $(\beta\rightarrow 0)$,
the SFF is highly sensitive to the differences in the energy levels of the system.
Interestingly, only the nearest neighbour energy spacings contribute to the very late time behavior. 
This enables the SFF to play a crucial role in understanding the time dynamics of 
chaotic quantum systems and also makes it a very useful tool for probing the discreteness 
in their energy spectrum.
A property of chaotic quantum systems is that they satisfy the \emph{Wigner's surmise}, which is a statement about the distribution of the nearest neighbour spacings of the energy levels. This qualifies the SFF as a suitable observable for analysing chaos in such quantum systems.

\noindent
As we learnt in the previous sections,
a lot is known about the energy spectrum of RMTs in the large $N$ limit. When the SFF is averaged over an 
ensemble of random matrices, it has a very particular behavior, especially at late times \cite{Cotler:2016fpe, Brezin:1977sv}.
In this light, we wish to understand the general dependence of the SFF on the specific nature of the RMTs 
and understand the time-scales for the onset of RMT behaviour in quantum chaotic systems. The idea is to consider the various non-Gaussian 
ensembles with polynomial potentials discussed in the previous section, and average the 
time-dependent SFF over all these ensembles. All such ensembles have a distribution of eigenvalues that are
different from each other, except at small energy scales. We wish to understand in what sense the SFF is 
sensitive to the difference in the ensembles, which would tell us its effectiveness in analysing the onset of quantum chaos.
The quantity we wish to study is defined as,
\begin{equation}
 G(\beta,t)=\f{\lan|Z(\beta+i t)|^2\ran_{\text{GUE}}}{\lan Z(\beta)\ran_{\text{GUE}}^2} \hspace{3.2cm}
\nn \\ =\f{\int\,d\lambda\, d\mu \,e^{-\beta(\lambda+\mu)}e^{-i t(\lambda-\mu)}
\big\lan\md(\lambda)\md(\mu)\big\ran_{\text{GUE}}}
{\int\,d\lambda\, d\mu \,e^{-\beta(\lambda+\mu)}
\big\lan\md(\lambda)\ran\lan\md(\mu)\big\ran_{\text{GUE}}}
\label{SFF}
\end{equation}
where $\md(\lambda)$ is the eigenvalue density and the integration is over the (unscaled) support of $\rho(\lambda)$ in each case.
For simplicity and clarity of the analysis, we wish to focus on the $\beta=0$ (or $T=\infty$) window. In the large $N$ limit,
the two point function of the level densities in (\ref{SFF}) can be divided into two parts;
the connected part ($G_c$) and the disconnected part ($G_{dc}$), which we define below.
\begin{equation}
 G_{dc}=\f{\lan Z(\beta + i t)\ran\lan Z(\beta - i t)\ran}{\lan Z(\beta) \ran^2}\bigg|_{\beta=0} \hspace{2cm} \nonumber \\
 \hspace{2cm}=\f{\int\,d\lambda \,d\mu \,e^{-i t(\lambda-\mu)}\big\lan\md(\lambda)\big\ran\big\lan\md(\mu)\big\ran}
 {\int\,d\lambda \,d\mu\big\lan\md(\lambda)\big\ran\big\lan\md(\mu)\big\ran}
 \label{SFF-dc}
 \end{equation}
 \begin{equation}
G_{c} = (G-G_{dc})|_{\beta=0} 
\label{SFF-con}
\end{equation}
We define the quantity \eqref{SFF-con} as the \emph{connected} two-point correlation function. In fact, it is easy to 
show that if we consider fluctuations around the extremised value of the level density,
\begin{equation}
\md(\lambda)=\bar{\md}(\lambda) + \delta \md(\lambda)\nonumber
\end{equation}
then the connected two point correlation function can be written as just $\lan\delta\md(\lambda)\delta\md(\mu)\ran$.
The connected part of SFF can then be re-written as,
\begin{equation}
G_{c}= \f{\int\,d\lambda\, d\mu \,e^{-i t(\lambda-\mu)}\lan\delta\md(\lambda)\delta\md(\mu)\ran}
{{\int\,d\lambda \,d\mu\big\lan\md(\lambda)\big\ran\big\lan\md(\mu)\big\ran}}
\label{SFF-c}
\end{equation}
We wish to analyse the behavior of these two parts separately in RMTs with \emph{non-Gaussian} potentials. We shall consider the quartic and sextic potentials and analyse the behavior of the
SFF, when averaged over these ensembles, using expressions for the level density we found in the previous sections.
\begin{center}
\begin{figure}[t]
 \centering
 \includegraphics[height=140 pt, width =220 pt]{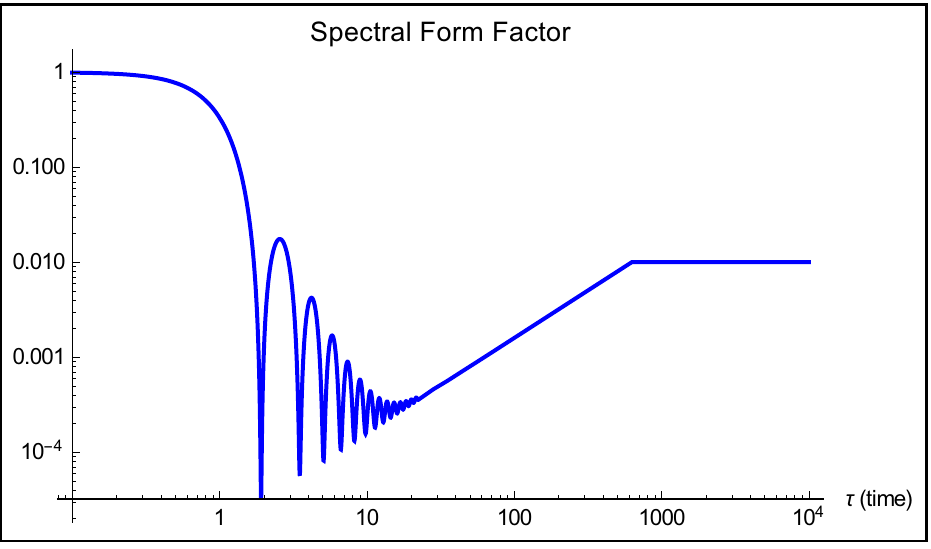}
 \caption{\footnotesize{The Spectral Form Factor for a GUE with N=100. The initial decay is due to the disconnected part of the SFF. The ramp and plateau is due to connected part of the SFF.  
 The estimated dip-time is $\tau\sim 10$. The ramp ranges from $\tau\sim\mathcal{O}(\sqrt{N})$ to $\tau\sim\mathcal{O}(N)$, beyond which the 
 plateau appears. The ramp and the plateau appear to be universal for all polynomial potentials $V(\l)$ due to the
 universality of the short distance correlators of the level densities. The initial decay, however, changes at the critical points
 of these polynomial potentials.}}
\label{fig-3}
\end{figure}
\end{center}
\subsection*{Convention}
For calculating the SFF, we shall use the mean level density in the one cut phase of the quartic model, given by (\ref{density-quartic}).
Before proceeding, we shall put in place some comments regarding the conventions we will be using, in order to connect up 
to results in the rest of the literature. The mean level density can be normalised in two ways. 
\begin{equation}
\int_{-2\alpha}^{2\alpha}\mathcal{D}(\lambda)\,d\lambda=N
\end{equation}
where $\mathcal{D}(\lambda)$ counts the total number of eigenvalues between $(\lambda,\lambda + d\lambda)$
and $\alpha\sim\mathcal{O}(\sqrt{N})$. This is also the expression we mean to use for 
calculating the SFF (since $\lan Z(\beta+i t)\ran|_{\beta,t=0}=N$). However, this is not the mean level density we get 
from extremising the action since all factors of $N$ have been scaled out from our action and all eigenvalues are just $\mathcal{O}(1)$.
The density we get from our action is normalized as,
\begin{equation}
\int_{-2a}^{2a}\rho(\lambda)\,d\lambda=1
\end{equation}
where the interval parameter $a\sim\mathcal{O}(1)$. The two are related by a $\sqrt{N}$ scaling of the eigenvalues.
If we rewrite $G_c$ and $G_{dc}$ in terms of this scaled density, all expressions would remain the same 
(with appropriate scalings for the coupling constants wherever necessary) except for two changes,
\begin{enumerate}
 \item The intergration limits would correspond to $a\sim\mathcal{O}(1)$.
 \item The time shall be scaled accordingly as $\tau=\sqrt{N}t$.
\end{enumerate}
Using the scaled mean level density we then proceed to calculate the SFF.
\subsection{The Disconnected Part : $G_{dc}$}
\subsection*{Quartic Model}
To calculate the disconnected part of the SFF, we use the scaled mean level density of the one-cut potential (\ref{density-quartic}) into the first term of (\ref{SFF-dc}).
\begin{equation}
\lan Z(\beta + i t)\ran_{\text{nGUE}}|_{\beta=0}=\int\,d\lambda  \,e^{-i \tau\lambda}
\big\lan\rho(\lambda)\big\ran_{\text{nGUE}}
 \nonumber \\ = \int_{-2a}^{2a}\,d\lambda \,e^{-i \tau\lambda} \f1{\pi}\bigg(\f12 + 2 g \lambda^2 + 4g a^2\bigg)\sqrt{4 a^2-\lambda^2}\nonumber\\
\nonumber \\= \f1{\tau^2}\bigg(a\, \tau\,(1 + 24 a^2 g )\, J_1(2 a \tau) - 24 a^2 g\, J_2(2 a \tau)\bigg)
\label{dc}
\end{equation}
The above expression \footnote{Note that the expression for SFF on the L.H.S. is in terms of $t$ 
but all expressions on the R.H.S. are in terms of $\tau$.} has a very intriguing form. 
The reason being that the coefficient of the first term 
vanishes exactly at the thermodynamic critical point, $g=g_c(=-1/48)$, of the quartic model, since
\begin{equation}
a^2=\f1{24g}(\sqrt{1+48g}-1)|_{g=g_c}=-\f1{24g_c}
\end{equation}
This suggests that it is only the Bessel function of the 
second kind that governs the fall off behavior of the SFF at the critical point. This has the non-trivial 
implication that the SFF has a \emph{faster power law decay} at the critical point 
as compared to any other value of $g$. To understand this better,
we look at the large time asymptotics of the above expression and find,
\begin{widetext}
\begin{equation}
\lan Z(\beta+i t)\ran_{\text{nGUE}}|_{\beta=0}\xrightarrow{t\rightarrow\infty}
-\sqrt{\f{a}{\pi}}\f{(1+24 a^2 g)}{\tau^{3/2}}\cos\left(\f{\pi}4 + 2 a \tau\right)
-\f1{\sqrt{a\pi}}\f{(1+152 a^2 g)}{\tau^{5/2}}\sin\left(\f{\pi}4 + 2 a \tau\right)
+\mathcal{O}\left(\f1{\tau^{7/2}}\right)
\end{equation}
\end{widetext}
This clearly suggests that for all values other than the critical value, the power law assumes a $\tau^{-3/2}$ behavior.
However, at $g=g_c$ the coefficient of the $\tau^{-3/2}$ term vanishes indicating a crossover to a 
$\tau^{-5/2}$ power law decay behavior.

\noindent
One way to explain this behavior is to look at the form of the level density $(\rho(\lambda))$ at the critical point.
Away from the critical point, the level density has the form (\ref{density-quartic}) such that it goes to zero near the $\lambda=2a$ edge 
of the interval as $(\lambda-2a)^{1/2}$. This behavior results in the $\tau^{-3}$ power law beahviour of the SFF.
\noindent
At the critical point $g_c$, the SFF displays the following behavior,
\begin{equation}
 \rho(\lambda)=\f1{12\pi a^2}(4 a^2-\lambda^2)^{3/2}
\label{crit-quartic}
\end{equation}
This implies that at the $\lambda=2a$ edge, the level density now goes to zero as $(\lambda-2a)^{3/2}$. 
 The change in the power law decay behavior can then be completely attributed to this particular change in behavior of the level density near the edge of the spectrum. One can check this behaviour to be in accord with the Paley-Weiner theorem.
\subsection*{Sextic Model}
For the sextic model, the mean level density in the one-cut phase is given by (\ref{density-sextic}).
We shall work with the 
expression for the level density in the one cut phase and predict all critical points in the theory accessible from this phase. 
Calculating the SFF in this model, we get,
\begin{widetext}
\begin{align}
&\lan Z(\beta+ i t)\ran_{\text{nGUE}}|_{\beta=0}=\int\,d\lambda  \,e^{-i \tau\lambda}
\big\lan\rho(\lambda)\big\ran_{\text{nGUE}}\nonumber\\
&= \int_{-2a}^{2a}\,d\lambda \,e^{-i \tau\lambda} \f1{\pi}\bigg(3h\lambda^4+(2g+6h a^2)\lambda^2+\bigg(\f12 + 4g a^2+18 h a^4\bigg)\bigg)\sqrt{4 a^2-\lambda^2}\nonumber\\
&= \f1{\tau^4}\bigg(a(-360 a^2 h \tau+ (1 + 24 a^2 g +180 a^4 h)\tau^3)\, J_1(2 a \tau) - 24 a^2(-30 h+(g +15 a^2 h)\tau^2 )J_2(2 a \tau)\bigg)
\label{dc2}
\end{align}
\end{widetext}
This expression is subject to a constraint equation,
\begin{equation}
 60\, h a^6 + 12 g a^4 +a^2 -1 =0
\label{constraint2}
\end{equation}

It is easy to check that putting $h=0$ in the above expression gives us back (\ref{dc}). To check that our method correctly predicts the critical points (as it did in the quartic model), we shall try to predict the critical points of the sextic model in a similar way. We shall look at the vanishing of the coefficients of the various powers of $\tau$ from the expression for the large time expansion of (\ref{dc2}).\\
The large time behavior of the SFF is,
\begin{widetext}
\begin{align}
\lan Z(\beta+i t)\ran_{\text{nGUE}}|_{\beta=0}&\xrightarrow{t\rightarrow\infty}
-\sqrt{\f{a}{\pi}}\f{(1+24 a^2 g+180 a^4 h)}{\tau^{3/2}}\cos\bigg(\f{\pi}4 + 2 a t\bigg)\nonumber\\
&+\f{3}{16}\f1{\sqrt{a\pi}}\f{(1+152 a^2 g+2100 a^4 h)}{\tau^{5/2}}\sin\bigg(\f{\pi}4 + 2 a \tau\bigg)\nonumber\\
&+\f{15}{512}\f1{a^{3/2}\sqrt{\pi}}\f{(-1+744 a^2 g+23628 a^4 h)}{\tau^{7/2}}\cos\bigg(\f{\pi}4 + 2 a \tau\bigg)
+\mathcal{O}\left(\f1{\tau^{7/2}}\right)
\label{large-t}
\end{align}
\end{widetext}

\begin{figure}[b]
 \includegraphics[width=220 pt, height =140pt]{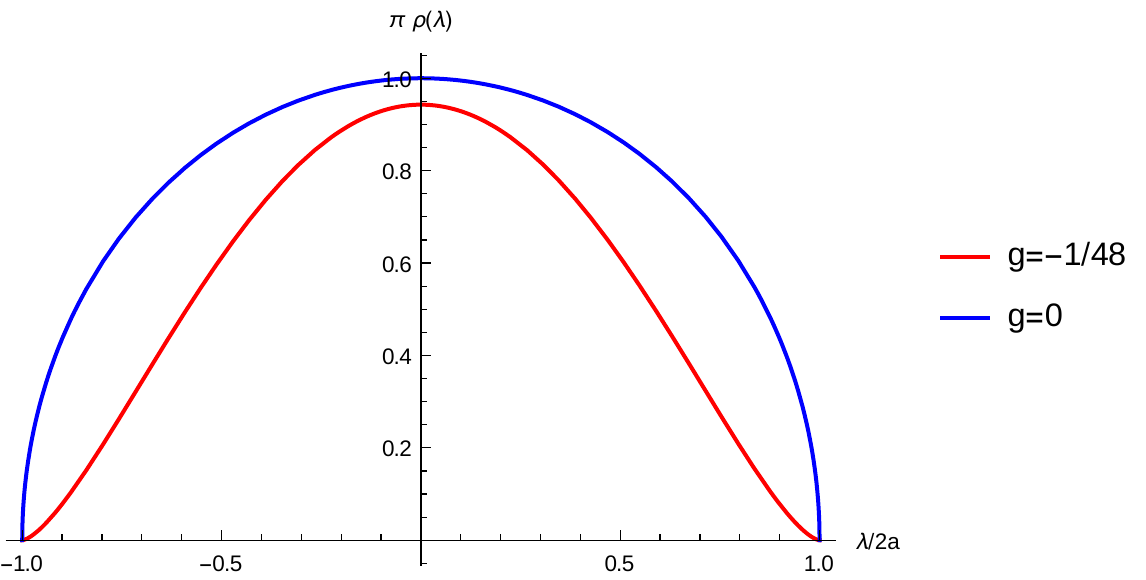}
  \caption{\footnotesize{ Plots for the edge behavior of the mean level density at the critical points in the quartic model.}}
\label{fig-41}
 \end{figure}
 
\noindent
For arbitrary values of $h$ and $g$, (\ref{large-t}) will have a large time fall off behavior governed by the $\tau^{-3/2}$ term.
However, to see the transition to a faster decay we would expect the coefficient of $\tau^{-3/2}$ to vanish while satisfying 
(\ref{constraint2}). Solving the set of simultaneous equations for $g$ and $h$, we find
\begin{equation}
 g= \f1{12 a^4}(3-2 a^2)\hspace{.5 cm}h= \f1{60 a^6}(a^2-2)
\label{crit-line}
\end{equation}
This is exactly the same as \eqref{g-h}. 
This, of course, is a one-parameter set of solutions. One has a solution for every value of $a\ge\sqrt{2}$. The lower bound is the value where,
$h=0$ and $g=-1/48$ which are the values for the critical point of the quartic model. The vanishing of this coefficient suggests that there is a 
transition of the power of the decay from $\tau^{-3/2}$ to $\tau^{-5/2}$.
The solution in terms of an unfixed parameter suggests that we 
have a \emph{critical line} separating the one and two cut phases of the sextic model. There is no upper bound on the 
value of $a$ yet. That can be obtained by analysing the point where the coefficient of the next highest power of 
$\tau$, namely $\tau^{-5/2}$, vanishes. Solving for the value of $a$ by plugging in the values of $g$ and $h$ from (\ref{crit-line}) into
the coefficient of the $\tau^{-5/2}$ term, we get $a=\sqrt{3}$ and correspondingly,
\begin{equation}
 g=-1/36\hspace{.3cm}\mbox{and}\hspace{.3cm}h=1/1620
\label{tri-crit}
\end{equation}
This is, again, the exact same point as in \eqref{tri-crit-pt}.
This indicates the place where the critical line, separating the one and two cut phases, encounters a \emph{tri-critical point}. It turns out that exactly at this point, the long time fall off behavior of the SFF changes over from $\tau^{-5/2}$ to 
$\tau^{-7/2}$.
 \begin{figure}[t]
  \includegraphics[width=220 pt, height =140pt]{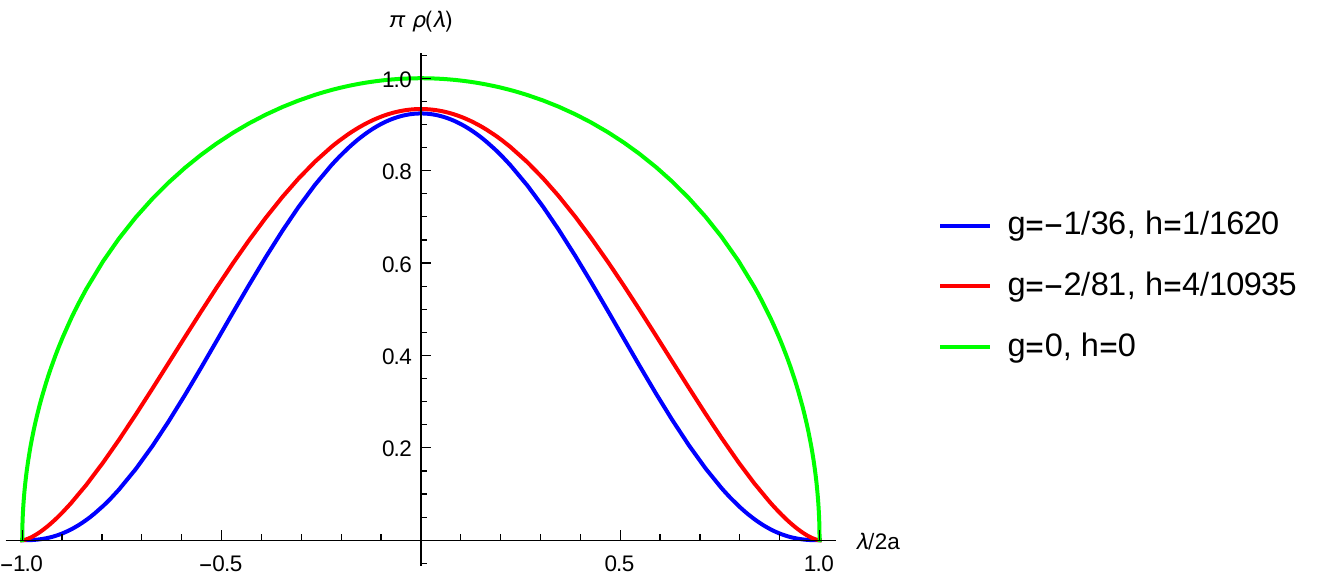}
 \caption{\footnotesize{ Plots for the edge behavior of the mean level density at the critical points in the sextic model.}}
\label{fig-42}
\end{figure}
\noindent
Again, this change in the fall-off behavior can be attributed to the change in the manner in which the mean level density approaches the
edge of one of its supports. Away from the critical point, $\rho(\lambda)\sim(\lambda-2a)^{1/2}$ near the $\lambda=2a$ edge.
Along the critical line and upto the tri-critical point,
\begin{equation}
 \rho(\lambda)=\f{(2 a^4 - 3 a^2 (\lambda^2-2)+6 \lambda^2)(4a^2-\lambda^2)^{3/2}}{60 \pi a^6}
\label{density-crit line}
\end{equation}
which implies that $\rho(\lambda)\sim(\lambda-2a)^{3/2}$ at the $\lambda=2a$ edge\footnote{The transition in the power-law behavior of the SFF, to a decay faster than $1/\tau^3$near criticality, as a result of the change in the edge behavior of the density of states was hinted at in \cite{Torres-Herrera:2018rde}. We would like to thank Antonio M. Garcia-Garcia for pointing this out to us. }.
Further, at the tri-critical point, the mean level density becomes
\begin{equation}
 \rho(\lambda)=\f1{540\pi}(\lambda^2-12)^{5/2}
\label{density-tri-crit}
\end{equation}
This clearly implies that $\rho(\lambda)\sim(\lambda-2a)^{5/2}$ at the $\lambda=2a$ edge. This then is the change responsible for the change in the 
behavior of the SFF from $\tau^{-5/2}$ to $\tau^{-7/2}$.
\subsection{Universality of the connected part: $G_{c}$}

\subsubsection{Universality of $G_c$ under the choice of potential}
The connected part of the SFF is as defined in (\ref{SFF-c}). It depends on the connected part of the two-point function of the mean level densities. In fact, the connected part depends only on the correlations of the fluctuations $\delta\rho(\lambda)$ around the mean-level densities. Let us use this fact to calculate some universal properties of $G_c$. Consider the continuum action (\ref{GUE}) but with a general potential $V(\lambda)$ (replace $\lambda^2$ with $2V(\lambda)$ in (\ref{GUE})). Let us calculate correlations of the fluctuations $\delta\rho(\lambda)$ around the mean-level density perturbatively.
Let  $ \rho(\lambda) = \rho_0 + \delta \rho(\lambda)$. Here $\rho_0$ is the large $N$ 
mean-level density. Plugging this into (\ref{GUE}) we get 
\begin{equation}
 \mathcal{Z} = \int \mathcal{D}\phi(\lambda)\ \exp\bigg[-\big( S_0 + S_1(\delta \rho) + S_2(\delta \rho^2)\big)\bigg]\nn
\end{equation}
where $S_0$ will be canceled by denominator in the definition of $G_{c}$ (\ref{SFF-c}), $S_1(\delta \rho)=0$ as $\rho_0$ is a large $N$ saddle-point solution, and
\begin{equation}
 S_2 = -N^2 \int d\lambda\ d\mu\ \delta \rho(\lambda)\ \log|\lambda - \mu|\ \delta \rho(\mu)
\end{equation}
Going over to Fourier space, we get
\begin{equation}
 S_2 = \frac{N^2}{2} \int d\tau\ \left(\delta \rho(\tau)\  \frac{1}{|\tau|}\ \delta \rho(-\tau) \right)
\end{equation}
With this expression, we identify (\ref{SFF-c}) to be $|\tau|/(2\pi N^2)$. This result then explains the \textit{ramp} in SFF. In order to analyse very large time behavior of the SFF, we take a long time average of SFF and get
\begin{equation}
 \lim_{T \to \infty} \int_0^T dt \left|\f{Z(\beta + i t)}{Z(\beta)}\right|^2 =  \f1{Z(\beta)^2}\sum_\lambda (g(\lambda))^2 e^{-2\beta \lambda}
\end{equation}
where $g(\l)$ is the degeneracy in $\l$. For a non-degenrate spectrum, $g(\l) =1 \ \forall\ \l$, which then gives 
\begin{equation}
  \lim_{T \to \infty} \int_0^T dt \left|\f{Z(\beta + i t)}{Z(\beta)}\right|^2 = \f{Z(2\beta)}{Z(\beta)^2}\hspace{0.2cm} \xrightarrow[]{\beta \to 0} \frac{1}{N}
\end{equation}
The RHS, in the above equation, is independent of time. This result then explains the origin of a time-independent \textit{plateau} in SFF.\\

\noindent \textbf{Note:}\textit{ The ramp and plateau behavior of the SFF is independent of the choice of potential. The connected part of the SFF displays a universal behavior under the choice of any polynomial potential, as long as the potential has the symmetries of the respective ensemble.}

\subsubsection{Universality of short distance correlations}
There is a more sophisticated way to understand this result.
A very standard result of RMT \cite{Mehta}
is the exact form of this correlator \emph{near the centre}
of the spectrum of the eigenvalues,
\begin{equation}
\lan\delta\rho(\lambda)\delta\rho(\mu)\ran = -\f{\sin^2[N(\lambda-\mu)]}{(\pi N(\lambda-\mu))^2}+\f1{\pi N}\delta(\lambda-\mu)
\label{sine} 
\end{equation}
This result is dubbed the \emph{sine kernel} and was originally derived using the method of Orthogonal polynomials
for the Gaussian ensembles.
However, we wish to emphasize that this result holds true for \emph{any} 
polynomial potential measure (of single trace operators) from which the ensemble of matrices is chosen. 
This point was previously mentioned in \cite{Brezin:1993qg}. In the context of the SYK model, it then suggests 
that all such polynomial potential models are equally good for describing its late time behavior.
A physical reason is that the various polynomial potentials significantly change the eigenvalue spectrum only 
near the edges(Fig.\ref{fig-21} and \ref{fig-22}). The correlations of the level densities near the center of the spectrum
($|\l|\sim0$), however, are observables that are oblivious to the change in the 
global structure of the eigenvalue density of the system, especially near the edges, and hence remain unchanged.
\noindent
\subsubsection{Fourier Transform}
To get to the SFF from (\ref{sine}), we need to plug it into the expression for the connected part in (\ref{SFF-c})\footnote{It should be noted that the scaling of $\mathcal{D}(\lambda)$ has been cancelled out between the numerator and the denominator, such that the same expression is valid with $\mathcal{D}(\lambda)$ replaced by $\rho({\lambda})$.}.
It is easy to see that with a change of variables from $(\lambda,\mu)$ to $(E=\lambda+\mu,\omega=\lambda-\mu)$, we can 
perform the integration. There are two parts to this integral, which we shall deal with separately.
\begin{enumerate}
 \item The $1/N^2$ part with the sine squared function that is responsible for the ramp. It is also the sub-dominant contribution.
 \item The $1/N$ part with the Dirac-delta function (the dominant part) which is reponsible for the plateau. 
\end{enumerate}

\subsubsection*{The Ramp and Plateau}
The integral involving the first term in (\ref{sine}) is responsible for the linear rise in time of the SFF, as mentioned earlier. We wish to show why this is the case. 
The expression for the SFF is given by,
\begin{equation}
 S(t)=\int_{-\infty}^{\infty} d\omega e^{-i \omega \tau}  \bigg[ -\bigg(\f{\sin(N\omega)}{N\omega \pi} \bigg)^2+\f{\delta(\omega)}{\pi N}\bigg]
\label{SFF3}
\end{equation}
where $\omega$ denotes the difference in energy levels. 
Evaluating the above integral gives us,
\[S(t)= \left\{\begin{array}{ll}
	  \f{\tau}{(2\pi N^2)}, & \tau<2N\hspace{.1cm}\mbox{or}\hspace{.1cm}t<2\sqrt{N}\vspace {.3cm}\\
	  \f1{\pi N} , &  \tau>2 N\hspace{.1cm}\mbox{or}\hspace{.1cm}t>2\sqrt{N}
        \end{array}\right.\label{ramp}\]
\noindent
These expressions tell us about the very long time behavior of the SFF at a fixed energy. At $\tau<2 N$, the SFF has a linear growth in
time. At $\tau\geq2 N$, this linear behavior saturates abruptly to a \emph{plateau} region.\\
From a physical point of view, it is enlightening to understand how this behavior arises. In (\ref{sine}), we can re-define the 
argument of the sine term in terms of,
\begin{equation}
 x= N (\lambda-\mu) = N\omega
\label{scaling}
\end{equation}
We would like to work in the limit where,
\begin{equation}
N\rightarrow\infty, \hspace{.2cm} \omega\rightarrow0 \hspace{.2cm}\mbox{and}\hspace{.2cm}x=\mbox{const.} 
\end{equation}
The scaling parameter $x$ in (\ref{scaling}) can be made small or large (by changing the way $\omega$ goes to zero) depending on the region of the spectrum we wish to focus on. 
\begin{enumerate}
 \item For large $x (\gg 1)$, $(\sin (x)/x)\sim 1/x$ in (\ref{sine}). 
 This behaviour implies that correlations between energy levels that are far apart, are suppressed as $1/N^2$. These fluctuations, in turn, do not contribute to the SFF. This behavior is dubbed the \emph{spectral rigidity}.  
 \item For small $x(\ll 1)$, $(\sin (x)/x)\rightarrow1$ in (\ref{sine}). This scaling focuses on the correlations between 
 fluctuations in the nearby energy levels. The integral in (\ref{SFF3}) gets its maximum contribution from near the $\omega=0$ 
 region. The non-zero value of the $(\sin (x)/x)$ part is responsible for the time dynamics of the SFF.
 These differences in the \emph{nearest} neighbour energy eigenvalues are therefore the ones that cause the ramp behavior of the SFF. 
 Evidently, the ramp stops when we reach the $\omega$ corresponding to values lower than the smallest eigenvalue spacing in the system, i.e. $\omega< \mathcal{O}(1/N)$.
\end{enumerate}
\section{Estimation of dip time\label{sec-4}}
One of the non-trivial effects of considering the SFF in a non-Gaussian RMT is the change in its initial fall-off behavior exactly at the critical points. This change also changes the dip-time for the various models we consider. As a corollary, this behaviour helps us uncover the timescale when the ramp starts, thereby giving us an upper bound on the ramp time \cite{Gharibyan:2018jrp}. The dip-time is estimated by comparing the initial fall-off behavior with the late-time behavior of the curve where it starts to rise linearly. We shall start by estimating the dip-time in the Gaussian model just to jog our memory, and quickly proceed to estimate it at the various critical points in the quartic and the sextic models.

\begin{figure}[t]
 \centering
 \includegraphics[height=140 pt, width =220 pt]{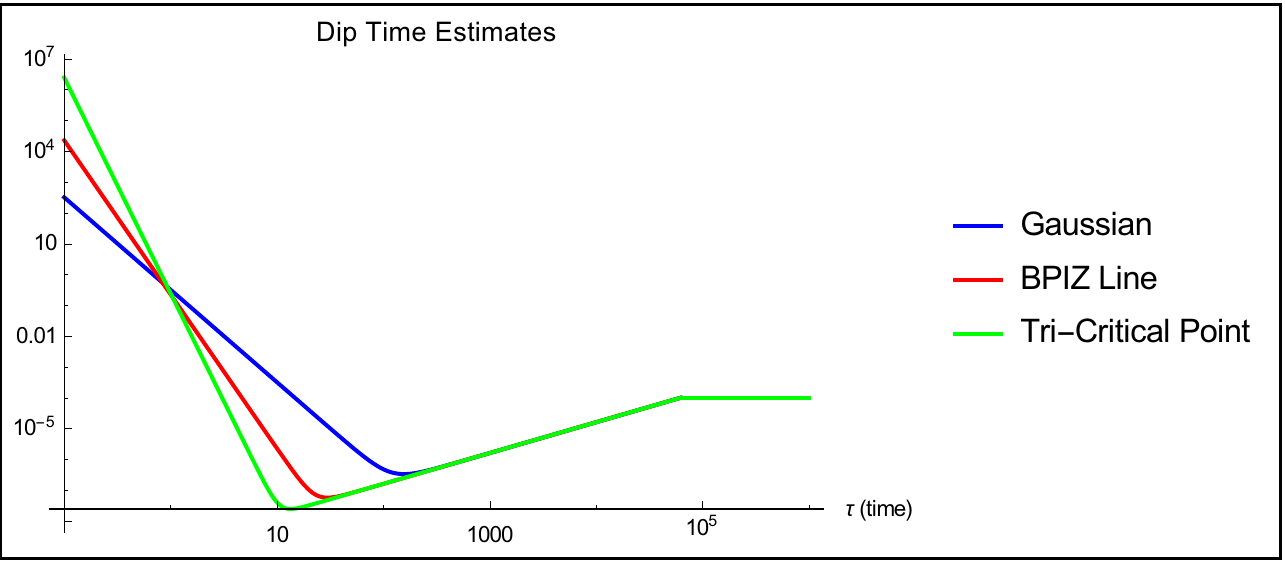}
 \caption{\footnotesize{The above figure shows the multicritical behavior of the SFF near the dip. The three graphs correspond to the 
 Gaussian ensemble ($g=0, h=0$), the BIPZ line ($g=-2/81, h=4/10935$) and the tri-critical point ($g=-1/36, h=1/1620$), respectively with N=10000.
 It shows the transition of the power law of the SFF at the different critical points.}}
\label{fig-5}
\end{figure}

\subsection{Gaussian RMT}
Equating the decay time from the disconnected part to the linear rise time from the connected part of the SFF in the Gaussian model, 
we find
\begin{equation}
 \tau^{-3}=\f{\tau}{N^2}\implies\tau\sim \mathcal{O}(\sqrt{N})\nonumber
\end{equation}
However, we should remember that $\tau=\sqrt{N} t$. Thus, in terms of the physical time parameter,
\begin{equation}
t\sim \mathcal{O}(1)\nonumber
 \end{equation}
which is the estimated dip-time for this model.
\subsection{Non-Gaussian RMT}
\subsection*{\underline{\textbf{Quartic Model}}}
\begin{enumerate}
 \item \underline{\textbf{Away from the critical point}}:
 Equating the decay rate with the linear rise, 
 away from the critical point of the quartic model, we see that
\begin{equation}
 \tau^{-3}=\f{\tau}{N^2}\implies\tau\sim \mathcal{O}(\sqrt{N})\nonumber
  \end{equation}
 which is the same as in the Gaussian model. Hence, the estimated dip time here would also be
\begin{equation}
 t\sim \mathcal{O}(1)\nonumber
  \end{equation}
 
 \item \underline{\textbf{At the Critical Point}}:
  Equating the decay time with the linear rise \emph{at} the critical point, we find
\begin{equation}
 \tau^{-5}=\f{\tau}{N^2}\implies\tau\sim \mathcal{O}(N^{1/3})\nonumber
  \end{equation}
 In terms of the physical time parameter, the estimated dip-time at the critical point is
\begin{equation}
 t\sim\mathcal{O}(N^{-1/6})\nonumber
  \end{equation}
\end{enumerate}

\subsection*{\underline{\textbf{Sextic Model}}}
\begin{enumerate}
 \item \underline{\textbf{Away from criticality}}: 
 Away from criticality, the estimated dip-time for this model is the same as that for the 
 previous models, namely 
\begin{equation}
 t\sim\mathcal{O}(1)\nonumber
  \end{equation}
 
 \item \underline{\textbf{On the critical line}}: 
 On the critical line for this model, we get
\begin{equation}
 \tau^{-5}=\f{\tau}{N^2}\implies\tau\sim \mathcal{O}(N^{1/3})\nonumber
  \end{equation}
 This behavior is the same as that at the critical point of the Quartic model. The 
 estimated dip-time in terms of the physical parameter is then,
\begin{equation}
 t\sim\mathcal{O}(N^{-1/6})\nonumber
  \end{equation}
 
 \item \underline{\textbf{At the tri-critical point}}: 
 At the tri-critical point,
 \begin{equation}
 \tau^{-7}=\f{\tau}{N^2}\implies\tau\sim \mathcal{O}(N^{1/4})\nonumber
  \end{equation}
 In terms of the physical time parameter, the estimated dip time is,
 \begin{equation}
 t\sim\mathcal{O}(N^{-1/4})\nonumber
 \end{equation}
\end{enumerate}

\section{Onset of Random Matrix Behaviour and Quantum Chaos}\label{sec-5}
The SFF has been developed as a tool for probing the nearest neighbour eigenvalue spacings in quantum systems. Due to the universality of the short distance correlations between eigenvalues, the very late time behavior of the SFF appears to be completely universal \cite{Mehta,Brezin:1993qg}. In Section III.B.1, we claim that this universality holds for all polynomial potentials $V(\lambda)$. In this section, we wish to point out the implications of this result on chaotic quantum systems via the SFF.

One very good example of a chaotic system is the SYK model. The SFF of this model, averaged over a Gaussian ensemble of the couplings, has been shown to behave like a Gaussian RMT at large times \cite{Cotler:2016fpe}. On another instance, the authors in \cite{Krajewski:2018lom} evaluated the large $N$ effective action for the SYK model averaged over a non-Gaussian ensemble of the couplings. The effect of such an averaging was found to shift the variance (proportional to $J$) by a term proportional to the non-Gaussian coupling. For the late time behaviour of the SFF, this would imply that the time-scale for the onset of ramp in the SFF will be affected (see Fig. 7 in \cite{Cotler:2016fpe}), without affecting the plateau in anyway. This is precisely our conclusion from the general analysis we carried out. To summarise our observations of universality in the late time behavior of the SFF averaged over arbitrary non-Gaussian ensembles with polynomial potentials,
\begin{enumerate}
 \item \emph{Any} RMT with a polynomial potential can describe a chaotic quantum system (including the SYK model) so long as 
 we concern ourselves with the low energy eigenvalues and their correlations. This behavior was also suggested
 in \cite{Verbaarschot:2000dy}, where RMT with arbitrary polynomial potential was shown to describe the IR behavior of QCD
 \footnote{We thank Nikhil Karthik for bringing this work to our attention}.
 \item The NNSD structure of the polynomial potential models seems to be similar to the Gaussian model, implying 
 that they too are well described by the \emph{Wigner's Surmise}. In the absence of analytic calculations, we shall
 present some numerical evidence for the suggested behavior in Appendix \ref{D}.
\end{enumerate}

\noindent
Another interesting aspect of the non-Gaussian averaging is the early time behaviour and actual onset of the ramp in chaotic systems. Away from criticality, the complete SFF does not, in any way, distinguish between the Gaussian and non-Gaussian ensembles. 
However, at the critical points, the early time decay of the SFF switches over to a faster fall-off behaviour. This change shows that the time scale at which ramp begins is much earlier than the dip time. In other words, it demonstrates a straightforward manner in which we can explore the actual ramp time in matrix models (see \cite{Gharibyan:2018jrp} for other ways to so). 

\section{Conclusions}
We have studied the Spectral Form Factor (SFF) in non-Gaussian random matrix models. We found the initial decay 
behavior of the SFF to be the same in both the Gaussian and non-Gaussian models \emph{except} at the 
critical points of the latter class, where it was found to display a faster decay.
We related this change in behavior to the change in the behavior of the mean eigenvalue density 
at the edges of its support, exactly at the critical points of the model. Moreover, at criticality, this faster decay helps us get a better upper bound on the \emph{ramp time}. This is important for the understanding of the onset of random matrix behaviour in a quantum chaotic system.

A crucial observation was made for the late time behavior of the SFF. Beyond the dip-time the SFF turned out to be universal
for any polynomial potential. This behavior was due to \emph{spectral rigidity}. With regard to implications on chaotic quantum systems, we emphasised 
that any random matrix models with a polynomial potential seems good enough to describe the late time chaotic behavior in such models. 

Due to this universal behavior of the SFF, one might think that the nearest neighbour spacing distribution in
all such models with polynomial potentials are well described by the \emph{Wigner-Dyson} distribution. We provided some numerical simulations to support this claim. However, analytic techniques to calculate the NNSD of the non-Gaussian models remains an open problem.
There are hints that this universal behavior persists for non-polynomial potentials as well. One interesting future direction would be to include double-trace operators and calculate the late-time behavior in such cases, to check if the aforesaid universality still persists. One very important goal of this project was to shed some light on how a chaotic quantum system displaying a typical scrambling behaviour at early times, gives way for the onset of a random matrix-like behaviour at late times. In this regard, we have been able to explore the various time-scales when RMT like behaviour might set into the system. However, in order to truly understand the cross-over, one would require an understanding of the phenomena of scrambling between the \emph{scrambling time} and the \emph{ramp time}, probably in a system with finite number of degrees of freedom. Scrambling on these time-scales would probably set the stage for the onset of an RMT description of the system beyond \emph{ramp time}.


In \cite{Dijkgraaf:2002fc, Dijkgraaf:2002dh}, a direct connection was made between matrix models (with non-Gaussian terms) and supersymmetric gauge theories, by identifying the superpotential of the gauge theory with the potential of the matrix model. One has to work with the gauged version of the matrix model to use this equivalence. However, it would be rather interesting to explore whether, in the spirit of \cite{Dijkgraaf:2002pp}, it is possible to directly explore the non-perturbative spectrum of the gauged SUSY theory from the perturbative dynamics of the matrix model, and thereby try to shed some light on their chaotic behavior.  
\begin{acknowledgements}
We thank Gautam Mandal for suggesting the problem and also for numerous important discussions during the course of this work. We also thank Saumen Datta, Sambuddha Sanyal, Geet Ghanshyam, Sounak Biswas and Sujoy Chakraborty for discussions on various aspects of the problem. Our computational work relied exclusively on the computational resources of the Department of Theoretical Physics at the Tata Institute of Fundamental Research (TIFR). R.S. would like to thank the Galileo Galilei Institute for Theoretical Physics for the hospitality and the INFN for partial support during the completion of this work. This work was also partly supported by Infosys Endowment for the study of the Quantum Structure of Space Time.
\end{acknowledgements}
\appendix
\section{Overview of RMT} \label{A}
Gaussian Matrix Ensembles are created by considering a large number of matrices, each of which are filled 
with random numbers drawn from a Gaussian distribution. Depending on whether the matrix elements are 
real, complex or quaternion numbers, we can have an orthogonal (GOE), unitary (GUE) or symplectic (GSE) ensemble
respectively. While describing real many-body systems, time reversal symmetric Hamiltonians can be described
by a GOE, whereas Hamiltonians with a broken time-reversal symmetry (for e.g. in a spin system with an external magnetic field) can be described by a GUE \cite{Mehta}.
The joint probabitlity distribution of such matrices is,
\begin{equation}
P(M)dM = \exp\left(-\f12\tr M^2\right)dM \hspace{3cm}\nonumber \\ =\exp \left(-\f12\sum_{i=1}^{N}x_{ii}^2\right)\exp \left(-\sum_{i\neq j}^{N}x_{ij}^2\right)\prod_{i\le j=1}^N dx_{ij}
\label{jpd}
\end{equation}
where $N$ is the rank of the matrix and the product in the measure is over all independent and identically distributed (i.i.d.)
variables.
\noindent
However, the L.H.S. of (\ref{jpd}) has a more generic meaning. It tells us that we can consider 
any ensemble of matrices that keeps the measure invariant under a similarity transformation,
\begin{equation}
M\rightarrow U^{-1}MU,\hspace{.1cm} \mbox{such that}\hspace{.1cm} P(U^{-1}MU)=P(M)
\end{equation}
where $U$ could be an orthogonal or a unitary matrix. This implies that the random elements 
can be drawn from any ensemble that satisfies the above condition. The most general ensemble for the time-independent
RMT can be written as,
\begin{equation}
\mathcal{Z} = \int dM\  e^{-Tr(V(M))}
\end{equation}
where $V(M)$ is the potential term.
The ensemble corresponding to the Gaussian weights is $V(M)=\f12 M^2$ . Let us work with this most general case and sketch out a 
solution in the large N limit along the lines of \cite{Brezin:1977sv}. \\
We start by diagonalizing the matrix via the above similarity transformation,
$M= U^{-1}D U$. The ensemble, written in the basis of the eigenvalues of the matrix, looks as follows,
\begin{align}
\mathcal{Z} &= \left(\prod_{i=1}^{N} \int d\lambda_i \right)
e^{-N\sum_{i=1}^{N} V(\lambda_i) + \beta \sum_{i<j}^{N} \log(\lambda_i-\lambda_j)}\nn\label{ensemble}\\
&=\left(\prod_{i=1}^{N} \int d\lambda_i \right) e^{-N^2 \left[S(\lambda_1,...\lambda_N)\right]}
\end{align}
where the action $S({\lambda_i})$ is,
\begin{equation}
S({\lambda_i}) = \frac{1}{N} \sum_{i=1}^{N} V(\lambda_i) -\beta \frac{1}{N^2} \sum_{i<j}^{N} \log|\lambda_i-\lambda_j|
\label{action}
\end{equation}
The first term in (\ref{action}) is just the potential written as a function of the eigenvalues. The additional $N$ has appeared in front of this
term as a result of the scaling of the eigenvalues by a factor of $\sqrt{N}$. This is done to ensure that the action doesn't scale 
with any factor of $N$. As a result, all eigenvalues are now $\mathcal{O}(1)$ numbers.
The second term is new and has appeared in the diagonalization process as a part of the measure. It is the \emph{Vandermonde determinant}
and it depends on the modulus of the difference between any pair of eigenvalues. It also depends on whether the matrices are  
orthogonal or unitary ($\beta =1 $ or $2$ accordingly). We shall only concern ourselves with GUE here and hence put $\beta=2$ in the above expression. To find a solution, we need to extremise the action w.r.t. $\lambda_i$, which gives us the 
following equation,
\begin{align}
\f{\partial S}{\partial \lambda_i} = 0 &\Rightarrow \f{1}{N} V'(\lambda_i) - \f{2}{N^2} \sum_{j\ne i} \frac{1}{\lambda_i-\lambda_j} = 0\nn\\
 V'(\lambda_i) &= \frac{2}{N} \sum_{j\ne i} \frac{1}{\lambda_i-\lambda_j}
 \label{eom}
\end{align}
The above integral equation was solved by \cite{Brezin:1977sv} using the method of resolvents that we shall describe below.


\section{The Method of Resolvents}\label{B}
Moving over to the continuum limit in the eigenvalues, we introduce a density of states $\rho(\lambda)$ containing information about 
the number of eigenvalues between $\lambda$ and $\lambda+ d\lambda$. In the continuum limit, the saddle point equation (\ref{eom}) becomes,
\begin{equation}
 V'(\lambda) = 2 Pr\bigg(\int\,d\mu \frac{\rho(\mu)}{\lambda-\mu}\bigg)
 \label{saddle-pt}
\end{equation}
where $Pr$ refers to only the principal part of the integral.\\
\noindent
The solution of this equation would give us the eigenvalue density (or level density) $\rho(\mu)$ which would
tell us about the distribution of the eigenvalues in the large $N$ limit. To solve equation (\ref{saddle-pt}), 
we shall use the \emph{method of resolvents} \cite{Brezin:1977sv}(see \cite{Eynard:2015aea} for a review)\footnote{To find other ways to calculate the density of states, see \cite{BREZIN1996697, Alexandrov:2006qx} and references within \cite{Eynard:2015aea}.}\footnote{This method only helps us find the leading large $N$ results. To get to the subleading $1/N$ corrections, see \cite{Migdal:1984gj, David:1990ge, KAZAKOV1991614, Ambjorn:1992gw}. We thank Andrei Mironov for pointing out these references to us.}.

\noindent
In the method of resolvents, we define the \emph{Stieltjes} transform $\omega(\lambda)$ for the function $\rho(\lambda)$ as,
\begin{equation}
\omega(\lambda)=\int_{\text{supp}\hspace{.1cm} \rho}\,d\mu \f{\rho(\mu)}{\lambda-\mu}
\end{equation}
For random matrices corresponding to real physical Hamiltonians, the eigenvalues are all real numbers. Thus, $\mu\in\mathbb{R}$.
Outside the support of $\rho(\lambda)$, the function $\omega(\lambda)$ has the folllowing asymptotic behavior\footnote{Assuming, $\int_{\text{supp} \mu}d\mu\, \rho(\mu)=1$},
\begin{equation}
\omega(\lambda\rightarrow\infty)=\f1\lambda + \mathcal{O}\left(\f1{\lambda^2}\right)
\end{equation}
The \emph{Stieltjes} transform extends the domain of the level density to complex values. Thus, $\lambda\in\mathbb{C}$.
It is then easy to check that, 
\begin{equation}
V'(\lambda)=\omega(\lambda+i 0)+\omega(\lambda-i 0), \hspace{.2cm} \lambda\in\mathbb{R}
\end{equation}
The property of the resolvent is also that,
\begin{equation}
\omega(\lambda+i 0)-\omega(\lambda-i 0)=-2\pi i \rho(\lambda)
\end{equation}
which means that it has a jump on the real line along the support of $\rho(\lambda)$.

Combining the above two properties, the resolvent assumes the following form,
\begin{equation}
\omega(\lambda\pm i 0)=\f12 V'(\lambda)\mp i\pi \rho(\lambda), \hspace{.2cm}\forall\ \lambda\in\text{supp}(\rho)
\end{equation}
The most general solution is,
\begin{equation}
\rho(\lambda)=\f1{2\pi}M(\lambda)\sqrt{-\sigma(\lambda)}
\end{equation}
where $M(\lambda)$ is some polynomial in $\lambda$ and $\sigma(\lambda)$ is a 
polynomial in $\lambda$ of the following form,
\begin{equation}
\sigma(\lambda)=\prod_{i=1}^{n}(\lambda-a_{2i-1})(\lambda-a_{2i})
\end{equation}
with $n$ being the no. of intervals on which $\rho(\lambda)$ is supported and 
$(a_{2i-1},a_{2i})$ being the end-points of the intervals.

\section{Level density for Polynomial Potentials}\label{C}
In case of the quartic potential, \cite{Brezin:1977sv} provided us with a solution to the integral equation using the resolvent method.
We wish to review that method here. The conditions satisfied by the resolvent are as follows,
\begin{enumerate}
 \item it is analytic in the entire complex plane $\mathbb{C}$ except on the interval $(-2a,2a)$ along the 
 real line.
 \item it is real everywhere outside the interval $(-2a,2a)$.
 \item it goes as $1/\lambda$ as $\lambda\rightarrow\infty$, with the eigenvalue density 
 appropriately normalised.
 \item Along the interval $(-2a,2a)$, the resolvent has a discontinuity of the following form,
 \begin{equation}
 \omega(\lambda + i 0)-\omega(\lambda- i 0)\hspace{3cm}\nn \\\hspace{3cm} =-2 i \pi \rho(\lambda)\hspace{.3cm}\lambda\in\mbox{supp }\rho
 \end{equation}
 \item From its definition, the resolvent satisfies the following property,
  \begin{equation}
 \omega(\lambda + i 0)+\omega(\lambda- i 0)=V'(\lambda)\nn \\
 V'(\lambda)=\lambda + 4 g \lambda ^3\ \lambda\in\mbox{supp }\rho
 \end{equation}
\end{enumerate}
\noindent
 The last two properties together suggest the following form for the resolvent,
   \begin{equation}
 \omega(\lambda\pm i 0)=\f12\lambda + 2 g \lambda ^3 \mp i \pi \rho(\lambda)
 \label{resolvent}
 \end{equation}
The solution for $\rho(\lambda)$ must conform to the analyticity properties for the resolvent
stated above. In fact, the unique form of the function that does so is,
 \begin{equation}
\rho(\lambda)=\f1{\pi}M(\lambda)\sqrt{-\sigma(\lambda)}
\end{equation}
where $M$ and $\sigma$ are polynomials in $\lambda$, with $\sigma(\lambda)$ of the following 
form,
 \begin{equation}
\sigma(\lambda)=\prod_{i=1}^{n}(\lambda-a_{2i-1})(\lambda-a_{2i})
\end{equation}
where $n$ is the no. of intervals on which $\rho(\lambda)$ is supported and 
$(a_{2i-1},a_{2i})$ are the end-points of the interval.\\
Suppose we have an even polynomial potential of degree 
$deg[V]=P$. 
Let $n$ be the no. of intervals on the real line  that form the
support of $\rho(\lambda)$. Then, for a solution of $\rho(\lambda)$, it 
can be seen that,
 \begin{equation}
deg[\sigma]=2n\hspace{.2cm}\mbox{and}\hspace{.2cm}deg[M]=P-1-n
\end{equation}
\subsubsection*{Quartic Potential}
For $P=4$ and $n=1$, this tells us that,
 \begin{equation}
deg[\sigma]=2\hspace{.2cm}\mbox{and}\hspace{.2cm}deg[M]=2
\label{1-cut-1}
\end{equation}
For $n=2$,
 \begin{equation}
deg[\sigma]=4\hspace{.2cm}\mbox{and}\hspace{.2cm}deg[M]=1
\end{equation}
However, the solution with two cuts in the quartic model is not physical, since it has complex solutions below
$g=g_c$. \\
Thus, the one-cut solution is the 
only one that can be determined in the quartic model. \\
\noindent
We assume the following form for $\rho(\lambda)$ in accord with (\ref{1-cut-1}), 
 \begin{equation}
\rho(\lambda)=\f1{\pi}(a_1 \lambda^2 + a_2)\sqrt{-(\lambda^2-4a^2)}
\end{equation}
Putting this into (\ref{resolvent}) (on the UHP) and expanding around $\lambda\rightarrow\infty$, we find
 \begin{equation}
(a_1,a_2)=\bigg(2g,\f12+4a^2 g\bigg)
\end{equation}
with the constraint,
 \begin{equation}
12 g a^4  + a^2 -1=0
\end{equation}
The form of the one-cut level density in the quartic case is,
 \begin{equation}
\rho(\lambda)=\f1{\pi}\bigg(\f12+4 a^2 g+ 2g\lambda^2\bigg)\sqrt{4a^2-\lambda^2}
\end{equation}
\subsubsection*{Sextic Potential}
For the sextic potential, $P=6$. Now we have three different possibilities
corresponding to the one, two and three-cuts. \\
\noindent
For $n=1$,
 \begin{equation}
deg[\sigma]=2\hspace{.2cm}\mbox{and}\hspace{.2cm}deg[M]=4
\label{1-cut-2}
\end{equation}
For $n=2$,
 \begin{equation}
deg[\sigma]=4\hspace{.2cm}\mbox{and}\hspace{.2cm}deg[M]=3
\end{equation}
For $n=3$,
 \begin{equation}
deg[\sigma]=6\hspace{.2cm}\mbox{and}\hspace{.2cm}deg[M]=2
\end{equation}
\noindent
For simplicity, we shall only display the method for finding the one-cut solution. The 
two and three cut solutions follow analogously.\\

\noindent
The unique form of the one-cut solution in accord with (\ref{1-cut-2}) is,
 \begin{equation}
\rho(\lambda)=\f1{\pi}(a_1 \lambda^4 + a_2\lambda^2+a_3)\sqrt{-(\lambda^2-4a^2)}
\end{equation}
Putting this form into the resolvent for the sextic model given by (\ref{sextic})
and expanding around $\lambda\rightarrow\infty$, we find that
 \begin{equation}
(a_1,a_2,a_3)=(3h, 2g + 6h a^2, 1/2 + 4 a^2 g + 18 a^4 h )
\end{equation}
with the constraint,
 \begin{equation}
 60 a^6 h+ 12 a^4 g + a^2-1=0 
\end{equation}
Thus, the unique form of the one-cut level density in the sextic case is,
\begin{widetext}
 \begin{equation}
\rho(\lambda)=\f1{\pi}\bigg(3h\lambda^4 + (2g + 6h a^2)\lambda^2 + \bigg(\f12 + 4 a^2 g + 18 a^4 h\bigg)\bigg)\sqrt{4a^2-\lambda^2}
\end{equation}
\end{widetext}

\section{NNSD in Non-Gaussian Models}\label{D}
\begin{figure}[H]
 \centering
 \includegraphics[height=100 pt, width =180 pt]{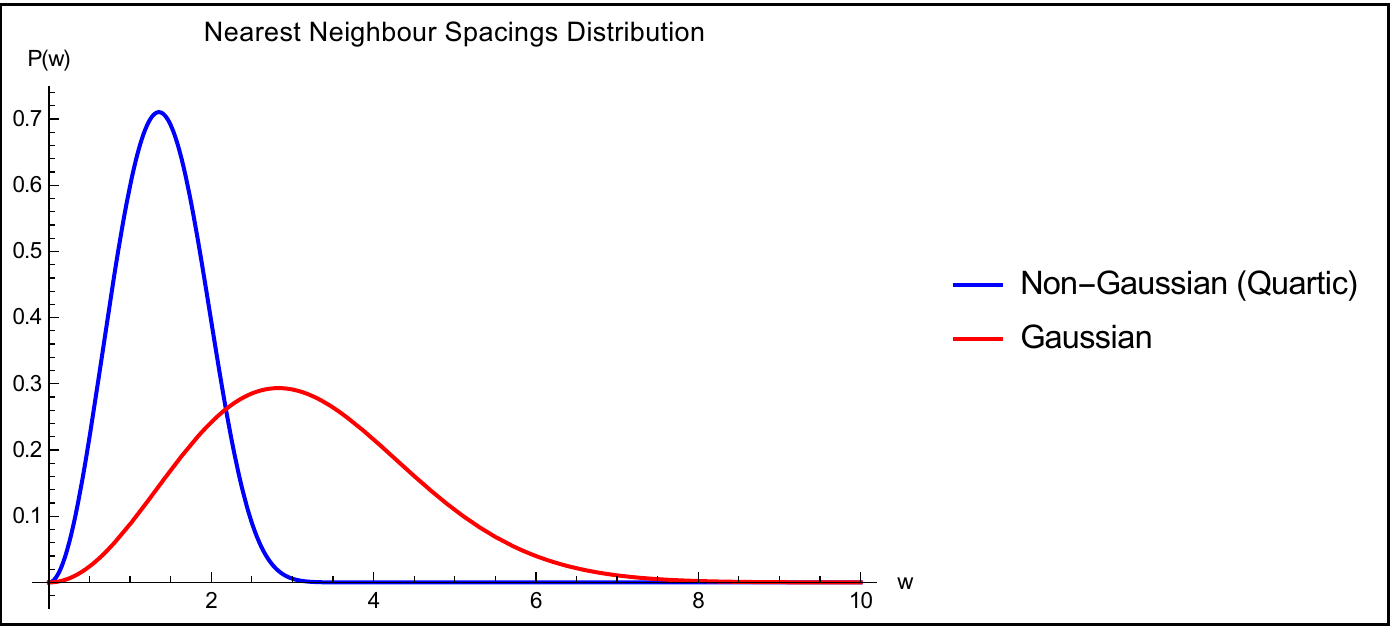}
 \caption{\footnotesize{This plot shows the difference in the NNSD between the Gaussian and non-Gaussian ensemble (quartic model with $g=1$).
 The most notable difference is the non-Gaussian tail of the non-Gaussian ensemble as $\omega\rightarrow\infty$.}}
\label{fig-6}
\end{figure}

The nearest neighbour spacing distribution (NNSD) has been the hallmark of quantum chaos for a wide range of 
quantum mechanical systems. It is easily calculated for a large ensemble of two-level systems, where 
the different copies provide instances of neighbouring eigenvalues with varying differences. 
In the eigenvalue basis, it can be written as,

\begin{widetext}
 \begin{equation}
P(\omega)\propto \int\,d\l_1\,d\l_2 \,\delta(\omega - (\l_1-\l_2))|\l_1-\l_2|^2 e^{-\f12(\l_1^2 + \l_2^2)}
\label{Wig-surm1}
\end{equation}
\end{widetext}
where $\omega$ is the difference in the neighbouring eigenvalues. With the delta function, the integrals are easily evaluated 
to give,
 \begin{equation}
P(\omega)=A\omega^2 \exp(-B \omega^2)
\label{Wig-surm2}
\end{equation}
where $A$ and $B$ are related by the normalization condition: $\int_{0}^{\infty}P(\omega)=1$. The function 
$P(\omega)\sim\omega^2$ near $\omega\rightarrow0$ and has a Gaussian tail for $\omega\rightarrow\infty$.  
We wish to find similar expressions for the non-Gaussian ensembles. For that, we introduce the quartic terms in 
(\ref{Wig-surm1}) and define the new NNSD as,

\begin{widetext}
 \begin{equation}
P(\omega)\propto \int\,d\l_1\,d\l_2 \,\delta(\omega - (\l_1-\l_2))|\l_1-\l_2|^2 e^{-\f12(\l_1^2 + \l_2^2)-g(\l_1^4 +\l_2^4)}
\end{equation}
Integrating this, we get,
 \begin{equation}
P(\omega)\propto \left(\f{1+3 g \omega^2}{2g}\right)^{\f1{2}} K_{\f14}\left[\f{(1+3 g \omega^2)^2 } {32g} \right] \exp\bigg[\f{(1+g \omega^2(2+7g\omega^2))}{32 g}\bigg]
\label{new-Wig-surm}
\end{equation}
\end{widetext}
where $K_{\f14}(x)$ is the modified Bessel function of the second kind.
To understand its relevance, one must look at Fig.\ref{fig-6} . Analytically, one can understand the distribution better by looking at the 
two limits of $\omega$,
\begin{eqnarray}
P(\omega)=\left\{ \begin{array}{ll}
 \f{\omega^2}{\sqrt{2g}}e^{\big(\f1{32g}K_{\f14}(\f1{32g})\big)}   & \omega\rightarrow0\vspace{.5cm}\\
  2\omega\sqrt{\f{2\pi}{3g}}\hspace{.2cm}e^{-\f18\omega^2-\f{g}{16}\omega^4} & \omega\rightarrow\infty\nn
\end{array} \right.
\label{new-Wig-surm2}
\end{eqnarray}
Near $\omega\rightarrow0$, the distribution behaves analogous to the NNSD from the Gaussian ensemble, $P(\omega)\sim\omega^2$.
However, in the other limit ($\omega\rightarrow\infty$), it has a non-Gaussian tail in contrast with its Gaussian counterpart. 
So, for $g\neq0$, we see a deviation of $P(\omega)$ from the Wigner-Dyson distribution\footnote{Note that this function has no special behaviour at $g_c=-1/48$.}. However, from the form of $P(\omega)$
in Fig.\ref{fig-6}, added to the fact that the SFF diplays a universal behavior due to the universal behavior of the 
short distance correlations, one might speculate that the NNSD structure is still such that the system is chaotic. 
Our speculation is also partly based on an RG analysis that was carried out in \cite{Brezin:1993qg}.
According to their argument, the universality of the short distance correlations as well as the chaotic nature of the system
has much to do with the presence of a trivial stable fixed point at the origin in the space of couplings.

\bibliography{biblio}
\bibliographystyle{apsrev4-1}
\end{document}